\documentclass[preprint,showpacs,groupedaddress,prl,balancelastpage,11pt,onecolumn]{revtex4}
\usepackage{amsfonts}
\usepackage{amsmath}
\usepackage{amssymb}
\usepackage{graphicx}
\usepackage{dcolumn}
\usepackage{bm}

\begin{document}

\title{Description of electron transport dynamics in molecular devices: A
time-dependent density functional theoretical approach in momentum space
makes it simple}
\author{Zhongyuan Zhou}
\author{Shih-I Chu}
\affiliation{Department of Chemistry, University of Kansas, Lawrence, KS 66045}

\begin{abstract}
We propose a first-principles time-dependent density functional theoretical
(TDDFT) approach in momentum ($\mathcal{P}$) space for quantitative study of
electron transport in molecular devices under arbitrary biases. In this
approach, the basic equation of motion is a time-dependent
integrodifferential equation obtained by Fourier transform of the
time-dependent Kohn-Sham (TDKS) equation in spatial coordinate ($\mathcal{R}$%
) space. It is formally exact and includes all the effects and information
of the electron transport in molecular devices. The electron wavefunction is
calculated by solving this equation in a closed finite $\mathcal{P}$-space
volume. This approach is free of self-energy function and memory term and
beyond the wide-band limit (WBL). The feasibility and power of the approach
are demonstrated by the calculation of current through one-dimensional (1D)
systems.
\end{abstract}

\date{\today }
\maketitle

Building molecular electronic devices\ using individual molecules is one of
the ultimate goals in nanotechnology \cite%
{Collier99,Chen99,Rueckes00,Tans98,Joachim00,Park02}. A typical molecular
device consists of a molecule coupled to two (or more) electrodes \cite%
{Datta05}. Electrons in the molecular device can go anywhere along the
electrodes during the electron transport and electron wavefunctions may
extend to infinity in spatial coordinate ($\mathcal{R}$) space. Thus,
accurate calculation of the electron wavefunctions in the $\mathcal{R}$\
space is an intractable task for the study of electron transport.

To resolve this problem one of the widely used methods is to separate the
molecule from the electrodes and treat it as an open system. More
specifically, the molecular device is partitioned into left (L), central
(C), and right (R) zones \cite%
{Taylor01,Larade01,Brandbyge02,Ke04-1,Kurth05,Stefanucci06,Stefanucci08}.
The C zone is chosen to include the molecule and some atomic layers of the
electrodes so that the Hamiltonian and electron density of the L and R zones
are accurately described by the equilibrium bulk ones before the bias
applied \cite{Taylor01}. To calculate the electron wavefunctions, one
separates the C zone from the L and R zones and treat the C zone as an open
system. The effect of the L and R zones (environment) on the C zone (open
system) is characterized by a self-energy function \cite%
{Taylor01,Larade01,Brandbyge02,Ke04-1,Kurth05,Stefanucci06,Stefanucci08,Maciejko06,Zheng07}%
. After taking this effect into account, an equation of motion for the C
zone is\ derived. Solving this equation one achieves the electron
wavefunctions for further calculation. This scheme has been extensively
applied to the calculation of steady-state current in molecular devices \cite%
{Ke04-1,Brandbyge02,Taylor01,Larade01,Taylor03,Taylor02,Kaun03}.

However, this scheme is not completely applicable when a time-dependent bias
is applied. In this case, the equation of motion of the C zone is derived
from the time-dependent Kohn-Sham (TDKS) equation \cite%
{Kurth05,Stefanucci06,Stefanucci08}. It not only contains a two-time
self-energy function but also includes a memory term \cite%
{Kurth05,Stefanucci08,Maciejko06,Zheng07}. The self-energy function is
related to a two-time Green function that can only be calculated under the
wide-band limit (WBL) \cite{Wingreen89}, an approximation that neglects the
energy dependence of the coupling between the molecule and electrodes \cite%
{Maciejko06}. Beyond the WBL, the two-time Green function is governed by a
double integral equation which is generally unsolvable\ \cite{Maciejko06}.
The memory term depends on the two-time self-energy function. To calculate
the memory term at time $t$, one has to recalculate and store the
self-energy function at all the past time steps before this time. As a
result, computer resources such as CPU time and random-access memory (RAM)
required in the calculation become increasingly and thus extremely large
with increase of number of time steps.

To overcome these difficulties, a computationally feasible scheme was
proposed based on directly solving TDKS equation \cite%
{Kurth05,Stefanucci06,Stefanucci08}. In this scheme, the TDKS equation is
discretized in the whole $\mathcal{R}$ space and converted to a matrix
equation first, then an equation of motion is derived from the matrix
equation for the C zone, and finally the equation of motion is solved to
calculate the electron wavefunctions. This scheme has been successfully
applied to the study of 1D model systems \cite%
{Kurth05,Stefanucci06,Stefanucci08}. The time-dependent currents achieved
tend to the steady-state currents obtained from the Landauer formula \cite%
{Kurth05,Stefanucci06} and Floquet method \cite{Stefanucci08} after a long
time, providing a benchmark for the currents through the 1D model systems.
The most striking characteristic of this pioneering scheme is that the
effect of the L and R zones on the C-zone electrons and the reflection of
the time-dependent wavefunction on the boundaries are completely taken good
care of by a transparent boundary condition. This scheme can indeed
eliminate the explicit dependence of the equation of motion on the two-time
self-energy function. However, the equation of motion still contains a
memory term used to represent the effect of the L and R zones and remove the
reflection of the electron wavefunction on the boundaries. As a consequence,
the computer resources required in the calculation also increase with the
number of time steps.

In fact, the difficulties encountered in the $\mathcal{R}$-space
calculations at least come from one of the memory term and self-energy
function used to characterize the effect of the L and R zones on the
electrons in the C zone. Since the electron wavefunction can extend to the
infinity in the molecular device during the transport, the memory term
or/and self-energy function can not be disregarded no matter how large the C
zone is chosen. Because of this reason, quantitative study of electron
transient transport in the $\mathcal{R}$ space is currently a computational
challenge for realistic molecular devices.

However, the momentum of electron is always finite and less than certain
maximum value $\mathbf{k}_{\max }$ in the molecular device. The probability
of electron with momentum greater than $\mathbf{k}_{\max }$ is negligible or
zero. Thus, in momentum ($\mathcal{P}$) space the electron wavefunction is
localized (e.g., the wavefunction of a free electron spreads out to the
whole space in the $\mathcal{R}$ space but localizes in a very small volume
depicted by a $\delta $ function in the $\mathcal{P}$ space) \cite{zhou01}
and can be calculated in a finite volume as long as the boundary of the
volume is set at the place with properly large momentum $\mathbf{k}_{\max }$%
. In this case, all the troublesome terms related to the open system in the $%
\mathcal{R}$ space, such as the self-energy function and memory term, will
be wiped out completely and the difficulties encountered in the $\mathcal{R}$%
-space calculations can be resolved in the $\mathcal{P}$ space. Based on
this idea, we propose in this work a first-principles time-dependent density
functional theoretical (TDDFT) approach in the $\mathcal{P}$ space for the
quantitative study of electron transient transport in realistic molecular
devices under arbitrary biases.

The molecular device is in an equilibrium state described by a unique
temperature and chemical potential when time $t<0$. The charge of the two
electrodes is perfectly balanced and no current flows through the device. At
time $t=0$ a bias of voltage $v_{b}\left( t\right) $ is applied to the
electrodes. This bias drives the molecular device out of equilibrium and
induces the current through the device.

In the $\mathcal{R}$ space, the time-dependent electron wavefunction $\psi (%
\mathbf{r},t)$ is governed by the TDKS equation within TDDFT (atomic units
are used throughout the paper)%
\begin{equation}
i\frac{d}{dt}\psi (\mathbf{r},t)=\left[ h_{0}\left( \mathbf{r}\right)
+v_{D}\left( \mathbf{r},t\right) \right] \psi (\mathbf{r},t),  \label{t-3}
\end{equation}%
where, $h_{0}\left( \mathbf{r}\right) $ is the unperturbed KS Hamiltonian of
the electron and $v_{D}\left( \mathbf{r},t\right) $ is the driving potential
characterizing the effect of bias on the electron. Within the partition
above, the unperturbed KS Hamiltonian can be written as $h_{0}\left( \mathbf{%
r}\right) =-\nabla ^{2}/2+v_{0}\left( \mathbf{r}\right) $, where, $%
v_{0}\left( \mathbf{r}\right) =v_{BK\alpha }\left( \mathbf{r}\right) $ with $%
\alpha =L$ and $R$ for the L and R zones and $v_{0}\left( \mathbf{r}\right)
=v_{ext}\left( \mathbf{r}\right) +v_{eff}(\mathbf{r},0)$ for the C zone.
Here, $v_{BK\alpha }\left( \mathbf{r}\right) $ is the bulk potential of the
electrodes, $v_{ext}(\mathbf{r})$ is the external potential related to the
interaction between the electron and nuclei of the system \cite{Zhou05-1},
and $v_{eff}(\mathbf{r},0)$ is the effective potential of the electron
comprising the Hartree potential $v_{H}\left( \mathbf{r},0\right) $ and
exchange-correlation potential $v_{xc}(\mathbf{r},0)$. For the metallic
electrodes considered here, the driving potential $v_{D}\left( \mathbf{r}%
,t\right) $ is represented by $v_{L}\left( t\right) $ and $v_{R}\left(
t\right) $ in the L and R zones, respectively, and $v_{L}\left( t\right)
-v_{R}\left( t\right) =v_{b}\left( t\right) $. These potentials undergo
uniform time-dependent shifts due to the screening effect and thus do not
change with spatial coordinates if $v_{b}\left( t\right) $ is slowly varying
during a typical time scale (less than a fs) \cite{Stefanucci06}. In the C
zone, the driving potential $v_{D}\left( \mathbf{r},t\right) $ is
represented by $v_{C}\left( \mathbf{r},t\right) $ and $v_{C}\left( \mathbf{r}%
,t\right) =v_{eff}(\mathbf{r},t)-v_{eff}(\mathbf{r},0)$ is the perturbing
potential induced by the field of bias, where $v_{eff}(\mathbf{r},t)$ is the
bias-induced effective potential including the Hartree potential $%
v_{H}\left( \mathbf{r},t\right) $ and exchange-correlation potential $v_{xc}(%
\mathbf{r},t)$ \cite{Zhou07-8}. The bias-induced effective potential is
associated with the response of the C-zone electrons to the field of bias.
It may change dramatically and nonlinearly with spatial coordinates and can
only be calculated self-consistently with the C-zone electron wavefunctions.
The procedure proposed here is nonperturbative and goes beyond the linear
response approximation \cite{Zhou07-8}.

The relation between the time-dependent $\mathcal{R}$-space electron
wavefunction $\psi (\mathbf{r},t)$ and $\mathcal{P}$-space electron
wavefunction $\phi (\mathbf{k},t)$ is given by the Fourier transform $\psi (%
\mathbf{r},t)=\left( 2\pi \right) ^{-3/2}\int d\mathbf{k}\phi (\mathbf{k}%
,t)\exp \left( i\mathbf{k\cdot r}\right) $. Applying this transform to Eq. (%
\ref{t-3}), we obtain the $\mathcal{P}$-space TDKS integrodifferential
equation

\begin{equation}
i\frac{\partial }{\partial t}\phi (\mathbf{k},t)=\int d\mathbf{k}^{\prime }%
\left[ H_{0}(\mathbf{k,k}^{\prime })+V_{D}(\mathbf{k,k}^{\prime },t)\right]
\phi (\mathbf{k}^{\prime },t),  \label{E3}
\end{equation}%
where, $H_{0}\left( \mathbf{k},\mathbf{k}^{\prime }\right) =k^{2}\delta
\left( \mathbf{k}-\mathbf{k}^{\prime }\right) /2+V_{0}\left( \mathbf{k,k}%
^{\prime }\right) $ is the $\mathcal{P}$-space unperturbed Hamiltonian, $%
V_{0}\left( \mathbf{k,k}^{\prime }\right) $ and $V_{D}(\mathbf{k,k}^{\prime
},t)$ are the $\mathcal{P}$-space potentials calculated from the $\mathcal{R}
$-space potentials $v_{0}\left( \mathbf{r}\right) $ and $v_{D}\left( \mathbf{%
r},t\right) $ by the transform $V_{\alpha }(\mathbf{k,k}^{\prime })=\left(
2\pi \right) ^{-3}\int d\mathbf{r}v_{\alpha }(\mathbf{r})\exp \left[ i\left( 
\mathbf{k}^{\prime }-\mathbf{k}\right) \cdot \mathbf{r}\right] $ with $%
\alpha =0$ and $D$, respectively.

To calculate the $\mathcal{P}$-space electron wavefunction, we solve Eq. (%
\ref{E3}) numerically in a finite 3D volume $[-k_{\alpha \max }\leq
k_{\alpha }\leq k_{\alpha \max }|\alpha =x,y,z]$. The integrals in Eq. (\ref%
{E3}) are discretized using the generalized Legendre-Gauss-Lobatto
pseudospectral method \cite{Canuto88,Chu05,Zhou05-1} on 3D grid points $%
[0,1,\cdots ,N_{\alpha }|\alpha =x,y,z]$ in the finite 3D volume. Since the $%
\mathcal{P}$-space electron wavefunction is localized the boundary condition
for solving Eq. (\ref{E3}) is simply that the electron wavefunction is zero
on the boundary. After discretization, Eq. (\ref{E3}) is converted to a
time-dependent matrix equation%
\begin{equation}
i\frac{\partial }{\partial t}\mathbf{\Phi }(t)=\left[ \mathbf{H}_{0}+\mathbf{%
V}_{D}(t)\right] \mathbf{\Phi }(t)\mathbf{,}  \label{E-6}
\end{equation}%
where, $\mathbf{\Phi }(t)$ is an $N_{G}$-dimensional vector with the
components being the $\mathcal{P}$-space electron wavefunction $\phi (%
\mathbf{k},t)$ on the 3D grid points, where $N_{G}=N_{x}\times N_{y}\times
N_{z}$, $\mathbf{H}_{0}$ and $\mathbf{V}_{D}(t)$ are $N_{G}\times N_{G}$
matrices with the matrix elements constructed by the unperturbed Hamiltonian 
$H_{0}(\mathbf{k,k}^{\prime })$ and potential $V_{D}(\mathbf{k,k}^{\prime
},t)$ on the 3D grid points together with the weights of quadrature,
respectively.

The $\mathcal{P}$-space electron wavefunctions at any time can be calculated
from Eq. (\ref{E-6}) if initial electron wavefunctions are given. At finite
temperature, the electrons populate on the single electron states of the
unperturbed system according to the Fermi distribution function. While in
the case of low temperature limit considered here, the electrons will occupy
all the single electron states from the lowest energy up to the Fermi
energy. Thus, the initial electron wavefunctions are the eigenfunctions of
the unperturbed system below the Fermi energy. The eigenvalue equation for
the eigenfunctions is obtained by setting $t=0$ in Eq. (\ref{E-6}) and
replacing the term on the left hand side of Eq. (\ref{E-6}) by $E\mathbf{%
\Phi }(0)$. Note that $\mathbf{V}_{D}(t)=0$ when $t=0$.

Applying the second-order split-operator method \cite{Chu05} to Eq. (\ref%
{E-6}), one obtains the propagation equation $\mathbf{\Phi }(t+\Delta t)=%
\mathbf{P}_{D}\left( \tau \right) \mathbf{P}_{0}\mathbf{P}_{D}\left( \tau
\right) \mathbf{\Phi }(t)$, where, $\Delta t$ is the time step size, $\tau
=t+\Delta t/2$, $\mathbf{P}_{0}=e^{-i\mathbf{H}_{0}\Delta t}$\ and $\mathbf{P%
}_{D}\left( \tau \right) =e^{-i\mathbf{V}_{D}(\tau )\Delta t/2}$ are the
propagators, and $\mathbf{V}_{D}(\tau )=\left[ \mathbf{V}_{D}\left( t\right)
+\mathbf{V}_{D}\left( t+\Delta t\right) \right] /2$. The propagators can be
calculated using the eigendecomposition scheme of matrix \cite{Strang98}. If
the uniform time step size ($\Delta t=$ const.) is employed, the propagator $%
\mathbf{P}_{0}$ only needs to be calculated once. In contrast, the
time-dependent propagator $\mathbf{P}_{D}\left( \tau \right) $ has to be
calculated at every time step.

Using the $\mathcal{P}$-space electron wavefunctions, the current through
the molecular device is calculated by 
\begin{eqnarray}
I\left( t\right) &=&-\frac{1}{\left( 2\pi \right) ^{3}}\sum_{j=occ}%
\int_{S_{C}}d\sigma \widehat{\mathbf{n}}\cdot \text{Re }\left[ \int d\mathbf{%
k}\phi _{j}^{\ast }(\mathbf{k},t)e^{-i\mathbf{k\cdot r}}\right.  \notag \\
&&\times \left. \int d\mathbf{k}^{\prime }\phi _{j}(\mathbf{k}^{\prime },t)%
\mathbf{k}^{\prime }e^{i\mathbf{k}^{\prime }\mathbf{\cdot r}}\right] ,
\label{E-7}
\end{eqnarray}%
where, $S_{C}$ is the surface enclosing the C zone, $\widehat{\mathbf{n}}$
is the unit vector perpendicular to the surface element $d\sigma $, $\phi
_{j}(\mathbf{k},t)$ is the $j$th electron wavefunction with the momentum $%
\mathbf{k}$, and the sum $j$ is over all the occupied electron states of the
C zone.

To demonstrate the feasibility of the proposed $\mathcal{P}$-space approach,
we first apply it to a one-dimensional (1D) system driven by a DC bias. The
potential of the unperturbed 1D system is zero everywhere in the whole $%
\mathcal{R}$ space. The C zone extends from $x=-6$ to $+6$ a.u. Initially,
all the single electron states are occupied up to the Fermi energy $\epsilon
_{F}=0.3$ a.u. At $t=0$, a DC bias $v_{b}$ is applied to the electrodes. We
choose $v_{L}=-v_{R}=v_{b}/2$, $k_{\max }=2.0$ a.u., and $\Delta t=0.1$ a.u.
The identical 1D problem has been considered in Ref. \cite%
{Kurth05,Stefanucci06} by using the transparent-boundary approach in the $%
\mathcal{R}$ space. In FIG. 1, we plot the current densities at $x=0$ as a
function of time for three biases. It is shown that a steady-state current
is achieved for each bias and approaches the steady-state current value
obtained from the Landauer formula \cite{Kurth05,Stefanucci06}. Furthermore,
the currents of the proposed $\mathcal{P}$-space approach are in very
excellent agreement with those of the $\mathcal{R}$-space
transparent-boundary approach \cite{Kurth05,Stefanucci06}. However, the
computational effort in the calculation with the $\mathcal{P}$-space
approach is considerably smaller than that with the $\mathcal{R}$-space
approach.

The second 1D system we consider is a double square barrier potential driven
by a time-dependent bias. The potential of the unperturbed system is $%
v(x)=0.5$ a.u. for $5\leq \left\vert x\right\vert \leq 6$ a.u. and zero
otherwise \cite{Kurth05,Stefanucci06}. The C zone extends from $x=-6$ to $+6$
a.u. A time-dependent bias, $v_{b}\left( t\right) =v_{b0}\left[ \theta
\left( t\right) -\theta \left( t-t_{0}\right) \right] $, which is turned on
at $t=0$ and turned off at $t=75$ a.u., is applied to the electrodes, where $%
\theta $ is the step function and $t_{0}=75$ a.u. We choose $\epsilon
_{F}=0.3$ a.u., $v_{L}=v_{b}\left( t\right) $, $v_{R}=0$, $k_{\max }=2.0$
a.u., and $\Delta t=0.1$ a.u. In FIG. 2 we display the current densities at $%
x=0$ as a function of time for four biases. It is shown that for each bias
the current oscillates after the bias is turned off and the amplitude is
proportional to the bias. It is also shown that the currents tend to zero
steady-state current quickly after turning off the bias and the transient
time for turning off a bias is much shorter than that for turning on the
bias \cite{Kurth05,Stefanucci06}. Moreover, the results of the $\mathcal{P}$%
-space approach are in very good agreement with those of the $\mathcal{R}$%
-space transparent-boundary approach \cite{Kurth05,Stefanucci06}. Again the
proposed $\mathcal{P}$-space approach reproduces the results of the $%
\mathcal{R}$-space transparent-boundary approach with considerably less
computational effort.

In summary, we propose a first-principles TDDFT approach in the $\mathcal{P}$
space for the study of electron transport dynamics in molecular devices
under the arbitrary biases. This approach is effective in theoretical
treatment and efficient and feasible in computation. In this approach, the
electron wavefunction is calculated by solving a time-dependent
integrodifferential equation in a finite $\mathcal{P}$-space volume. This
equation is obtained by the Fourier transform of the $\mathcal{R}$-space
TDKS equation. It is exact and contains all the effects and information of
the electron transport of molecular devices. It is free of the tricky
self-energy function and memory term and beyond the extensively used WBL. In
addition, unlike the $\mathcal{R}$-space approach, the procedure based on
this approach does not need to impose any complicated boundary condition to
the electron wavefunctions. The computer resources such as CPU time used at
each time step and the RAM required in the calculation with this approach do
not increase with the number of time steps. The proposed $\mathcal{P}$-space
approach has been successfully applied to the study of 1D systems with less
computational effort, demonstrating it promising to extend the proposed
approach to the study of electron transport dynamics in realistic molecular
devices

\begin{acknowledgments}
ZZ is grateful to Dr. Gianluca Stefanucci for helpful discussions on the $%
\mathcal{R}$-space transparent-boundary approach. This work is partially
supported by the Chemical Sciences, Geosciences and Biosciences Division of
the Office of Basic Energy Sciences, Office of Science, U. S. Department of
Energy, and by the National Science Foundation.
\end{acknowledgments}

\bibliographystyle{apsrev}
\bibliography{nano1-prl}

\newpage

Figure captions

FIG. 1 (Color online) Current densities at $x=0$ for the biases $v_{b}=0.1$, 
$0.3$, and $0.5$ a.u. For each bias, the short dash curve is the result of
the $\mathcal{R}$-space transparent-boundary approach \cite%
{Kurth05,Stefanucci06}, the dash line is the steady-state current from the
Landauer formula \cite{Kurth05,Stefanucci06}, and the solid curve is the
result of the $\mathcal{P}$-space approach.

FIG.2 (Color online) Current densities at $x=0$ for the biases with $%
v_{b0}=0.05$, $0.15$, $0.25$ and $0.35$ a.u. For each bias, the short dash
curve is the result of the $\mathcal{R}$-space transparent-boundary approach 
\cite{Kurth05,Stefanucci06} and the solid curve is the result of the $%
\mathcal{P}$-space approach.

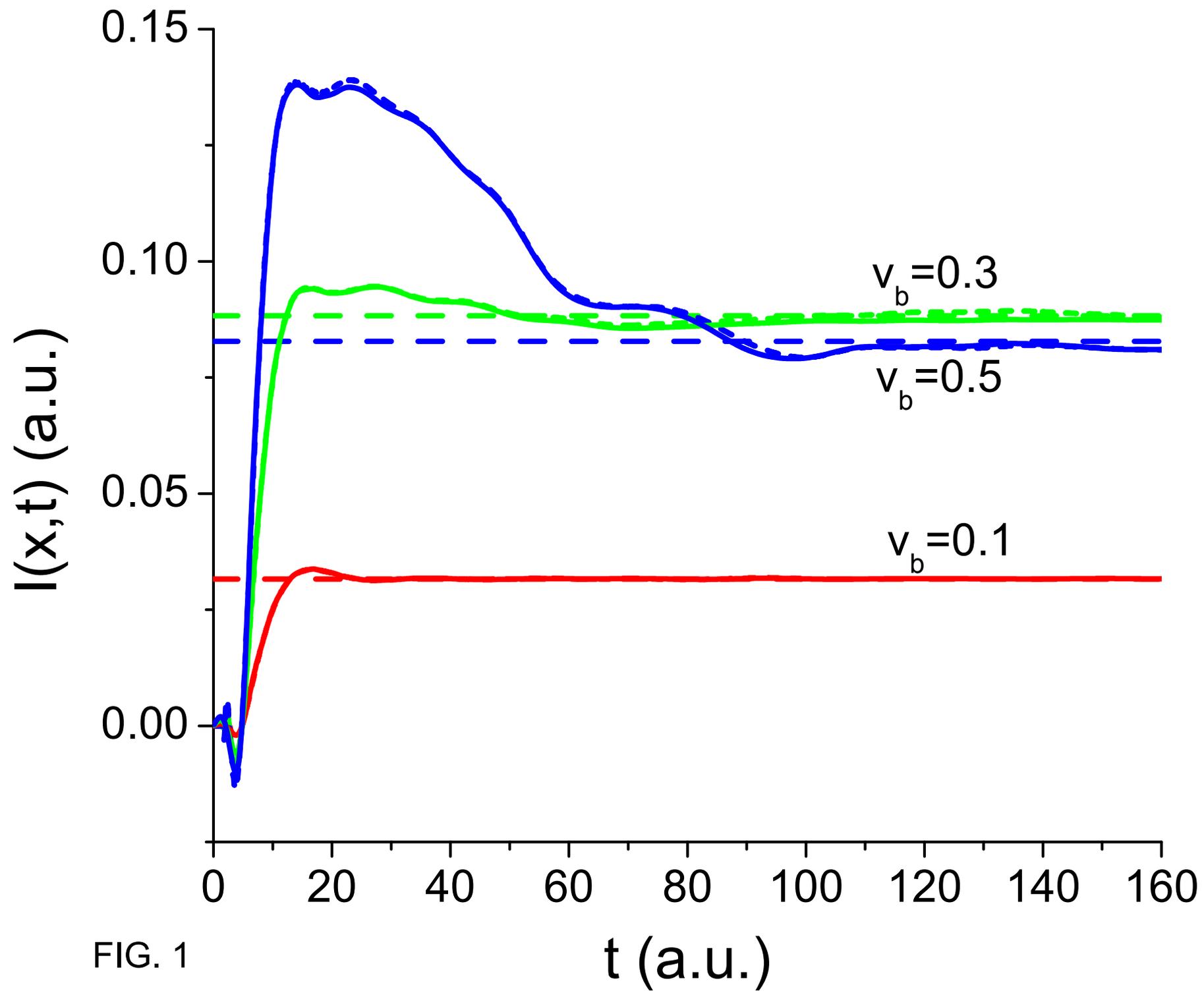
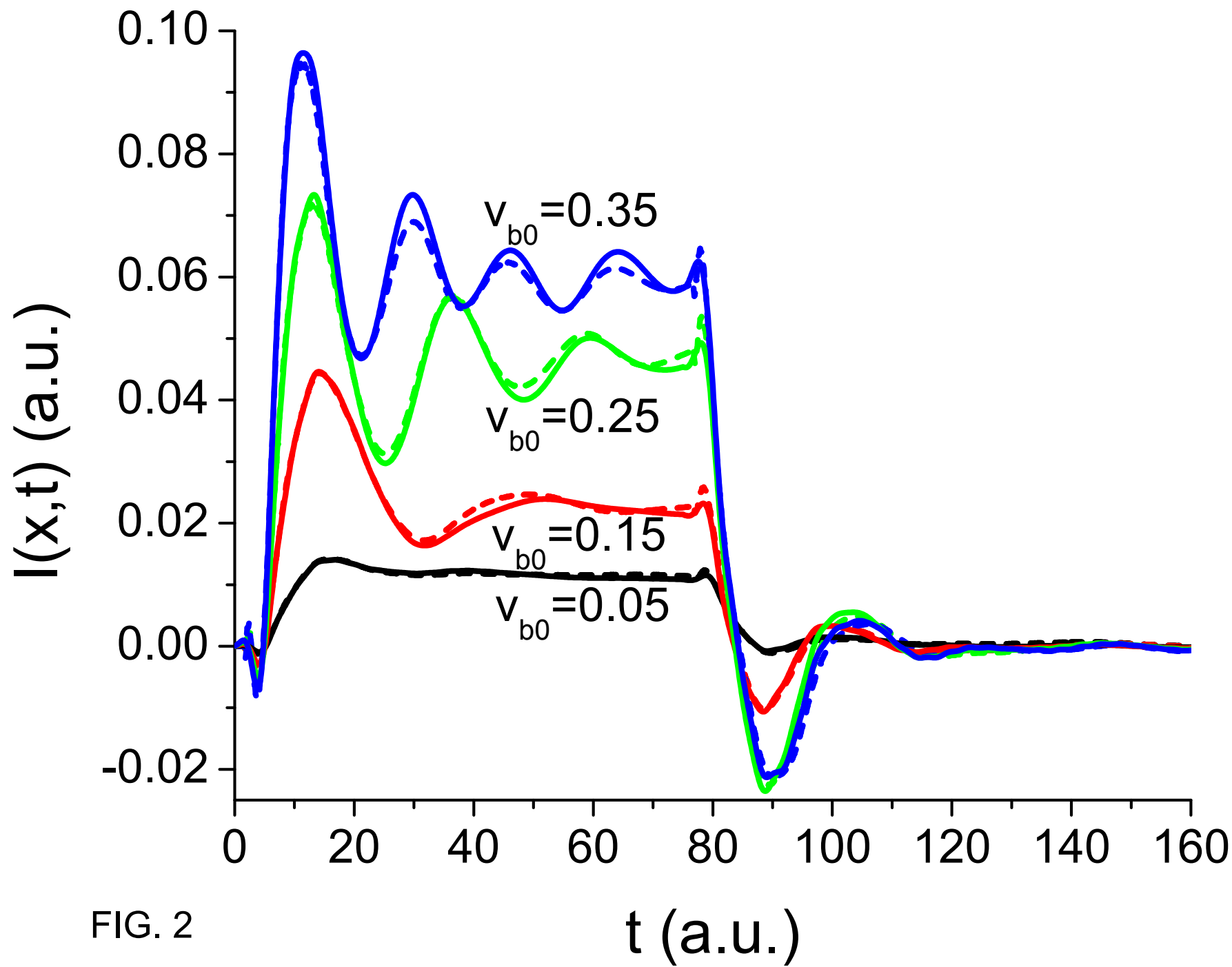
\end{document}